\documentclass{revtex4}

\usepackage{graphicx}
\newcommand{\ee}{ $\rm e^{+} e^{-}$ }

\newcommand{\ww}{ $\rm W^{+} W^{-}$}

\begin{document}

\bibliographystyle{revtex}


\begin{flushright}
LLR 02-001
\end{flushright}

\title{\boldmath The calorimetry   at the future \ee linear collider}




\author{Jean-Claude Brient}
\email[]{brient@poly.in2p3.fr}
\author{Henri Videau}
\email[]{Henry.Videau@poly.in2p3.fr}
\affiliation{Laboratoire Leprince-Ringuet - Ecole Polytechnique}

\date{\today}

\begin{abstract}
\end{abstract}

\maketitle

\section{Introduction}

The physics programme for  an electron-positron linear collider
 with center of mass energy ranging from
 90 GeV to 800/1000 GeV, is largely dominated by
 events with  final state containing many jets \cite{bib:jets}
 (essentially di-boson events with  W,Z or Higgs decaying into jets).\
 For an experiment on such an accelerator,
  the goal of the calorimeter can be either 
    to measure well the total flow of energy (in jets) or
 to measure the photons, the neutral hadrons and 
 to identify the leptons e, $\mu$ or $\tau$. The
  second option seems more difficult, but
our contention is that its three items can be optimised together 
and that, by that time, the energy flow measurement is optimised far
  beyond the first solution capability.
 This is done  through a so-called
energy flow algorithm which employs not only the calorimeter 
but also the tracker. The studies described below use a full
 full simulation of the calorimeter 
 named Mokka \cite{bib:mokka} based on GEANT4. 

\section{Physics programme and calorimeter performance}

First we can review the impact of the jet resolution on the physics programme:
This point has been looked at for few reactions up to now and needs a lot 
more work to assess it properly.

Parametrising the jet energy resolution as
 $\rm \Delta  E_{jet} = \alpha \sqrt{ E_{jet}}$,  two 
possibilities have been used, one with $\alpha$ = 0.6 corresponds
to a slightly improved LEP calorimeter, 
 the other, with $\alpha$ = 0.3,  corresponds
to what has been obtained in a full simulation and reconstruction with 
the Si-W electromagnetic calorimeter and the digital hadron calorimeter
described in the TESLA TDR \cite{bib:TDR}.

 First, the reaction 
\mbox{\ee $\to$Z H H}   which appears as 6 jet events.
  For an integrated luminosity
of 1 ab$^{-1}$,  the signal is observed at 3 sigmas for 0.6 and 
 at 6 sigmas for 0.3,
the difference between missing or getting it \cite{bib:TDR}.

\begin{figure}[ht]
\begin{center}
\mbox{
\includegraphics[height=7.0cm]{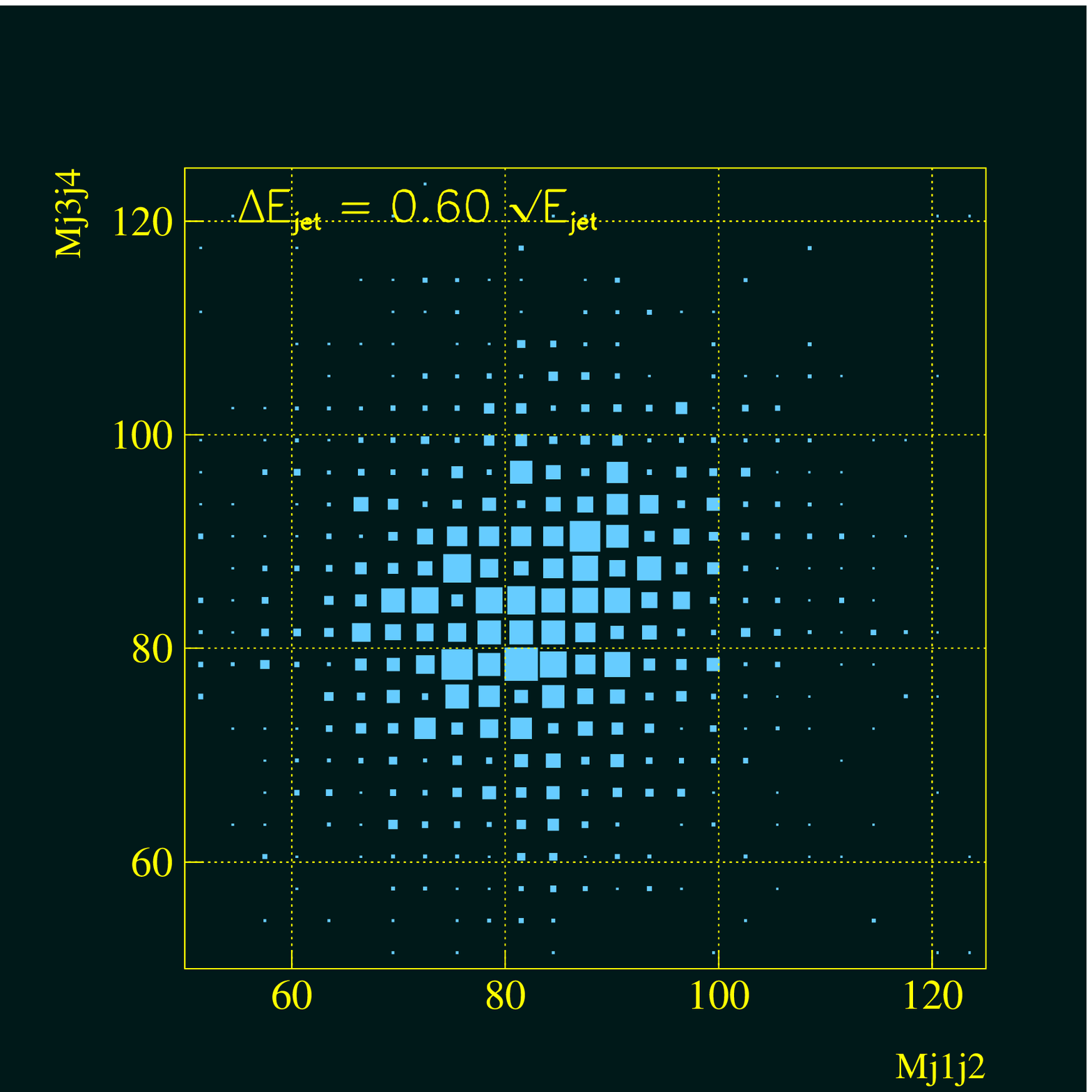}
\hspace*{1cm}
\includegraphics[height=7.0cm]{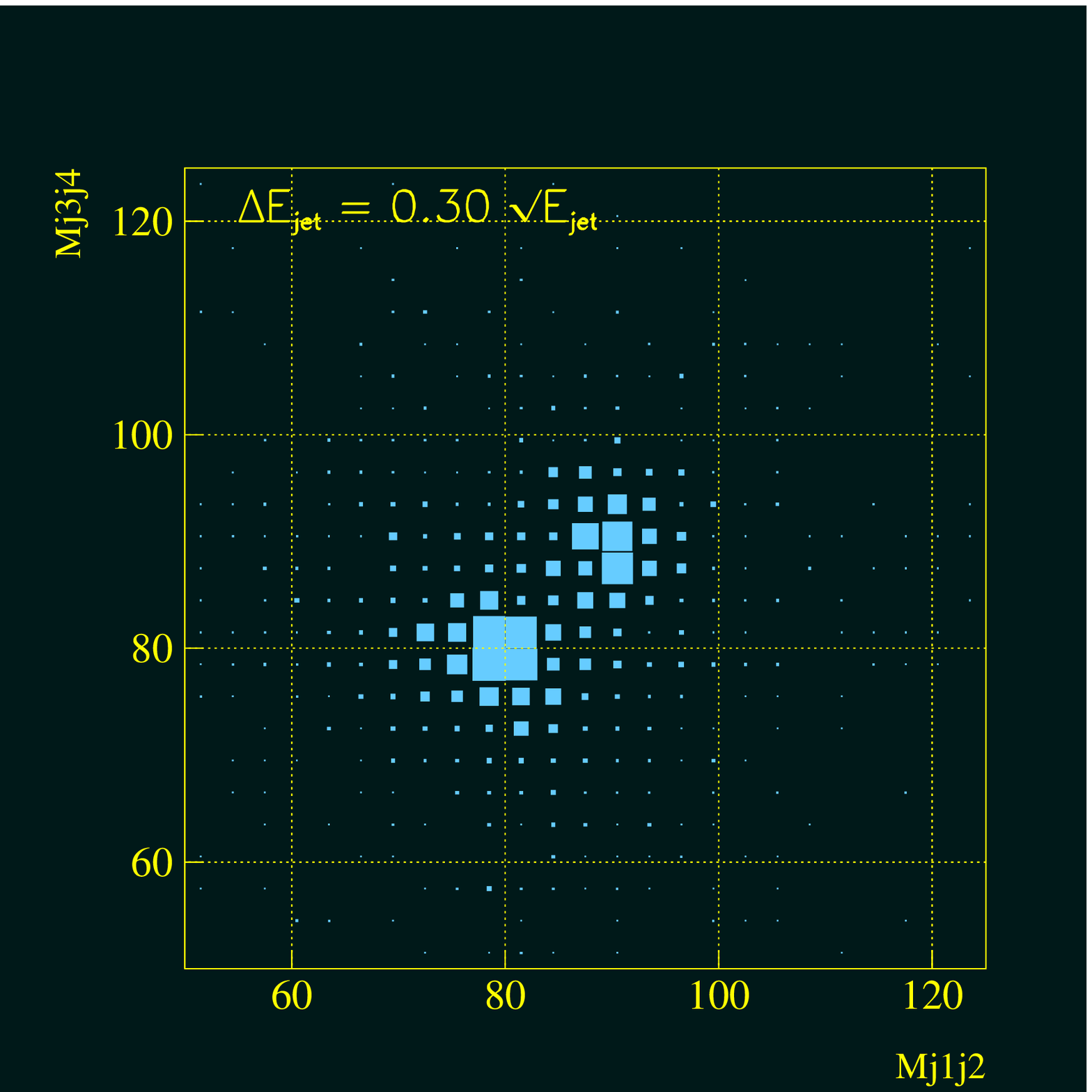}
}
\end{center}
\caption{ Impact of the jet  energy resolution
on the WW/ZZ separation, when $\alpha$ (see text)
 goes from 0.6 (left) to 0.3 (right). }
\label{fig:wwzz}
\end{figure}

A way to measure the impact of the resolution is to consider the amount
of statistics, running time, needed to obtain the same significance.
In the case of \ee$\to$ZH with Z into $q \bar{q}$ and H into WW$^*$, 
going from 0.3 to 0.6 
corresponds to loosing 45 \% of luminosity \cite{bib:jcbhenww}.
 In the same way,  
when separating ZZ$\nu \bar{\nu}$ from WW$\nu \bar{\nu}$
the loss is of 40 \% of luminosity.
This is illustrated in figure [\ref{fig:wwzz}].
The impact on the measurement of the process 
\ee $ \to t \bar{t} H$ is under consideration.
Considering the cost of running and the impact on physics of diluted runs
this is impressive. It has nevertheless to be studied with full simulation
and reconstruction and extended to many more reactions.

\section{The ECAL design}

Now that the physics case is painted, we can recall the basics of the 
analytical energy flow.
The energy flow of a jet, or a system of jets, is written as the sum of its 
components 
 \mbox{P=P$\rm _{Ch.particles}$ + P$_{\gamma}$ + P$\rm_{h^0}$}
($\rm h^0$ for neutral hadrons).
The argument is as follows: the charged hadrons make about 60\% of the jet
energy and, being of rather low energy, are much better measured in the 
tracker than in the hadron calorimeter. This point has to be tempered
when taking into account the track reconstruction efficiency, the generation
of fake tracks, the decays of particles like K$^{0}_{s}$.
Such a method relies more on separating properly the particles than on the
intrinsic energy resolution.
To get it, the calorimeter has to be far enough from the interaction but 
inside the coil, has to have a small radiation length, a small interaction 
length and a matched read-out granularity. Compact and granular.
There is one implicit parameter, the ratio of radiation to interaction lengths.
To measure properly the photons and to identify the electrons we need to
separate the electromagnetic and the hadronic primary components.
 This is
currently achieved up to a certain level longitudinally by going for a 
large ratio of radiation(L$_{X0}$) over interaction lengths
($\lambda_{I}$). We can compare the values
for the iron, the tungsten and the lead.
\begin{center}
\begin{tabular} {|c|r|c|r|}          \hline
 material & $\lambda_{I}$ (cm)  & L$_{X0}$  (cm)  & $\lambda_{I}$/  L$_{X0}$ \\ \hline
 Fe & 16.8 & 1.76 & 9.5 \\ \hline
 W  &  9.6      &   0.35   &   27.4 \\ \hline
 Pb &   17.1    &    0.56  &    30.5 \\ \hline
\end{tabular}
\end{center}

\begin{figure}[ht]
\begin{center}
\mbox{
\includegraphics[width=8.0cm]{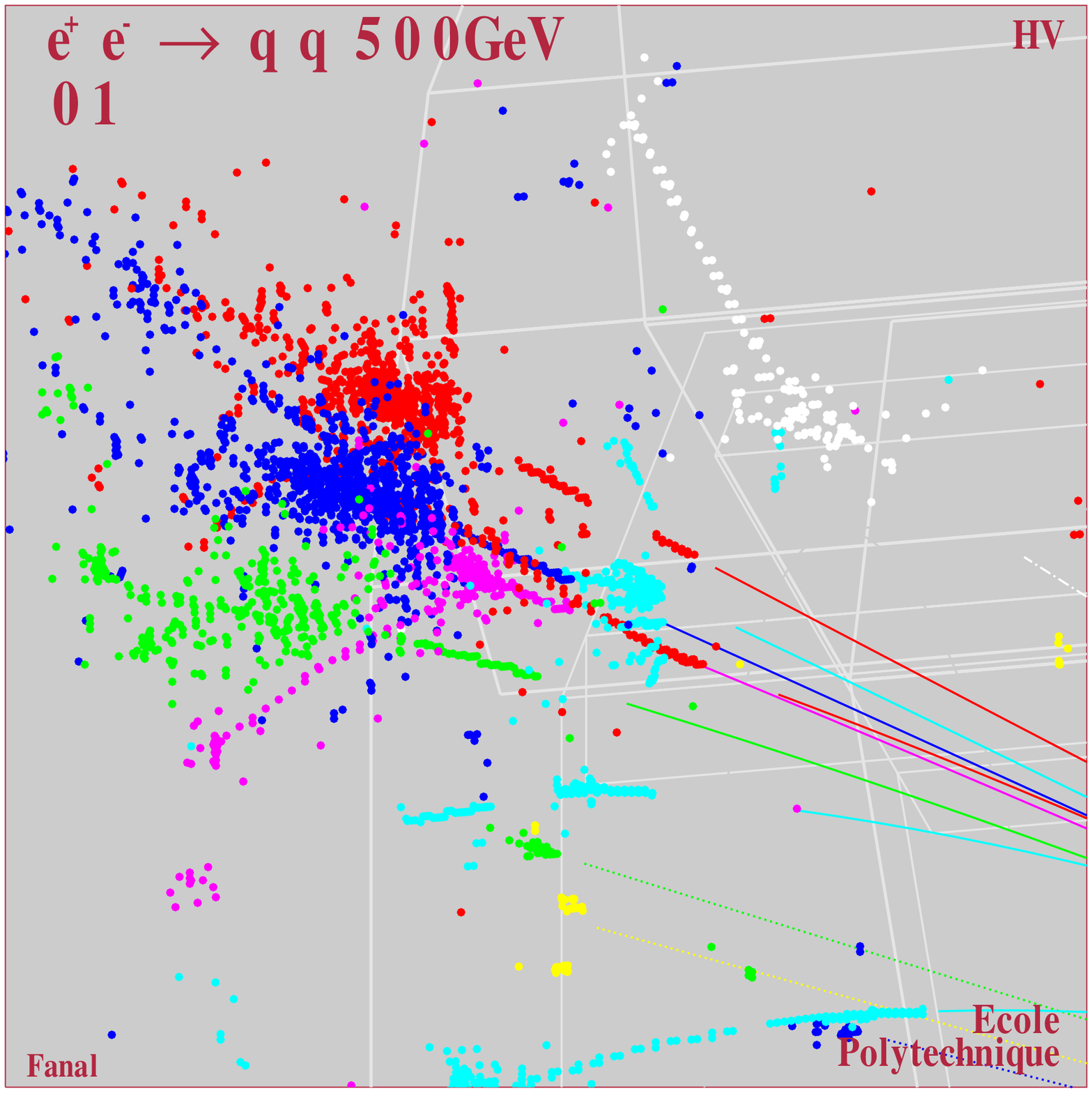}
\includegraphics[width=8.0cm]{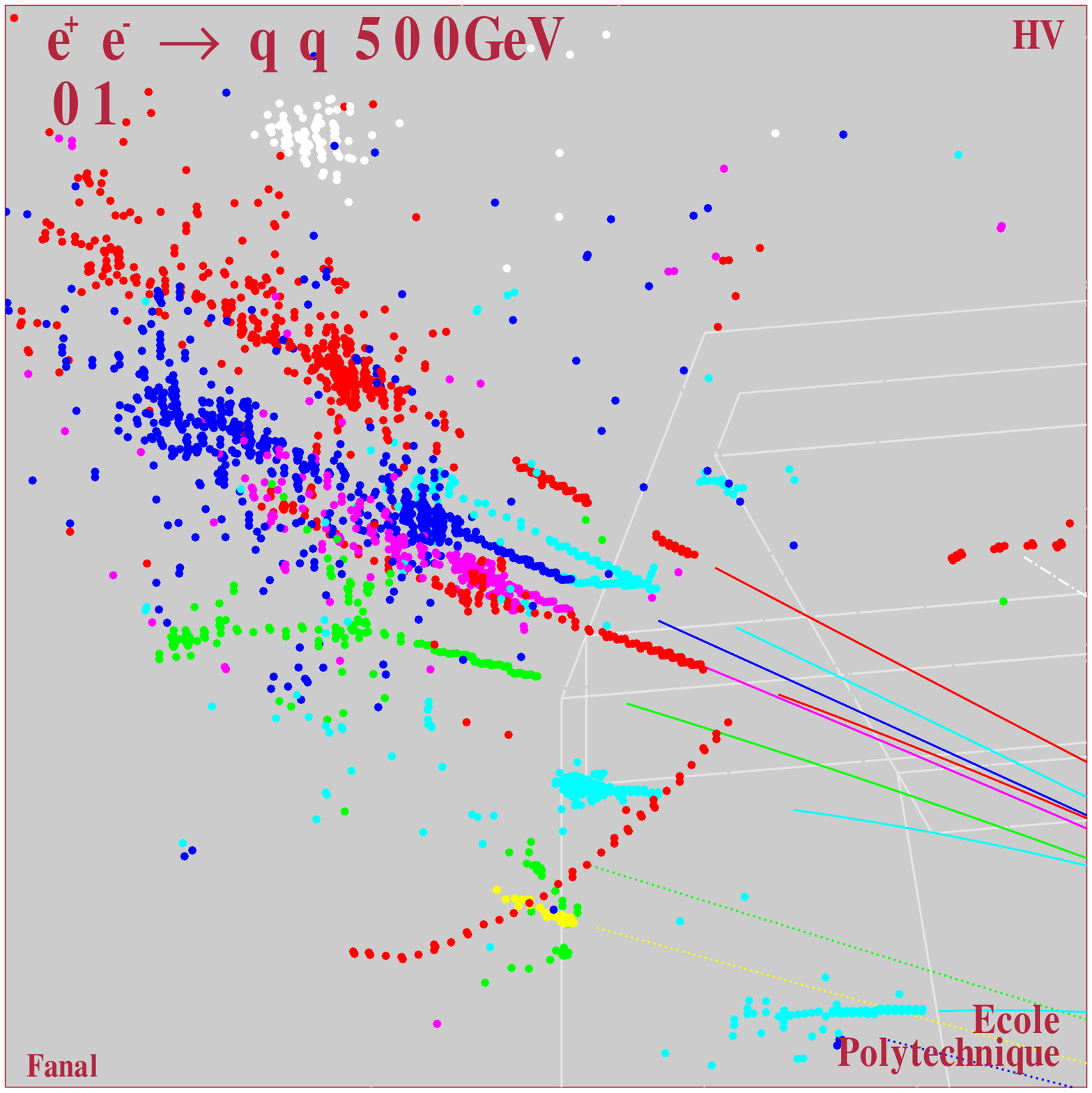}
}
\caption{ The same di-jet for a digital-steel (left)
 or for a digital-tungsten (right) HCAL}
\label{fig:tung-exp}
\end{center}
\end{figure}

The need of containing properly enough high energy electromagnetic showers
drives us to about 24 X0. This is not extremely sensitive since the 
longitudinal size grows logarithmically with energy and that we still get
information from the hadronic calorimeter behind.
Clearly we desire an electromagnetic calorimeter with a tungsten radiator
and a very compact detecting medium providing the adequate granularity,
a size close to the Molière radius. The only  solution known by the authors
 is  silicon diodes.
A clear obstacle seems to be, more than the technical difficulty, a question
of money. This will be discussed more thoroughly later.

The final parameters of the electromagnetic part we are left with are then:

- The radiator sampling or equivalently the number of layers. This drives
the intrinsic energy resolution and concurrently the price. We have considered
40 layers to reach a resolution  close to 0.1, but, if we really care only of
jet resolution, 20 may be enough as presented later.

- The cell size which provides the separation and the number of channels. It
has to be noted that a huge number of channels is a real technical challenge
but has little impact on the price, driven by the silicon area.
Currently we consider 1 to 1.5 cm$^2$.

\section{The HCAL design}

The same type of arguments, number of interaction lengths
 inside the coil,
the shower size, leads to compactness and favour tungsten. Reasons of cost,
of mechanical structure, and  may be some lack of reflection, 
have lead to choosing for the current design 
stainless steel for radiator, the eddy currents generated by a magnet quench
prohibiting copper. The structure is then 40 layers of 2cm thick iron plates
equivalent to 4 $\lambda_{I}$.
In fact this has to be worked out again playing with the
 $\lambda_{I}$/L$_{X0}$ ratio and with the thickness.
A hadronic shower can be seen as hadronic tracks connecting photon showers
($\pi^0$). Playing with  $\lambda_{I}$/  L$_{X0}$ 
will change the occupancy of the electromagnetic
subshowers with respect to the global hadronic shower. 
Reducing $\lambda_{I}$ will
enhance the ability to separate the showers, but, for a given interaction 
length sampling, will degrade the resolution.
An example is shown on figure \ref{fig:tung-exp}
 where the same jet is seen in iron and in
"expanded tungsten". This is tungsten where the density has been modified to
get the same fraction of interaction length in a plate as in the standard
iron plate.

We need now to focus on the hadronic cell structure.
There are two approaches, one, global, will match the cells to the hadronic 
shower width, the second tries to take advantage of the tree structure of
the shower and plays the same game as the electromagnetic part, trying
to isolate the electromagnetic component from the hadronic tracks.
In the first method the energy collected in a cell has to be measured,
 in the second you would just basically continue the ECAL, the cost precludes
such a solution but it happens that, for a proper granularity, a simple 
counting of the cells provides linearity of the response and an adequate
energy resolution.

These two solutions correspond grossly to the two variants of the TESLA TDR,
the variant with scintillator tiles varying in size between 5 and 20 cm,
\cite{bib.morgunov}, and the so-called digital variant with 1cm$^2$
cells.

\begin{figure}[ht]
\begin{center}
\includegraphics[width=8.0cm]{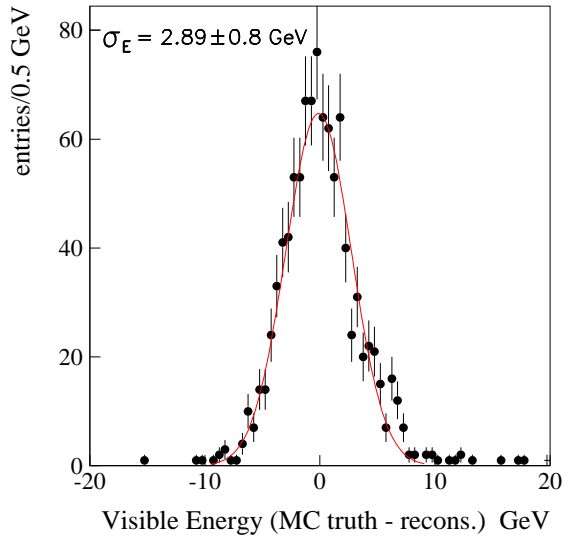}
\caption{The visible mass resolution for $\rm q \bar{q}$ Z decays at rest.}
\label{fig:Zmass}
\end{center}
\end{figure}
\begin{figure}[ht]
\begin{center}
\includegraphics[width=8.0cm]{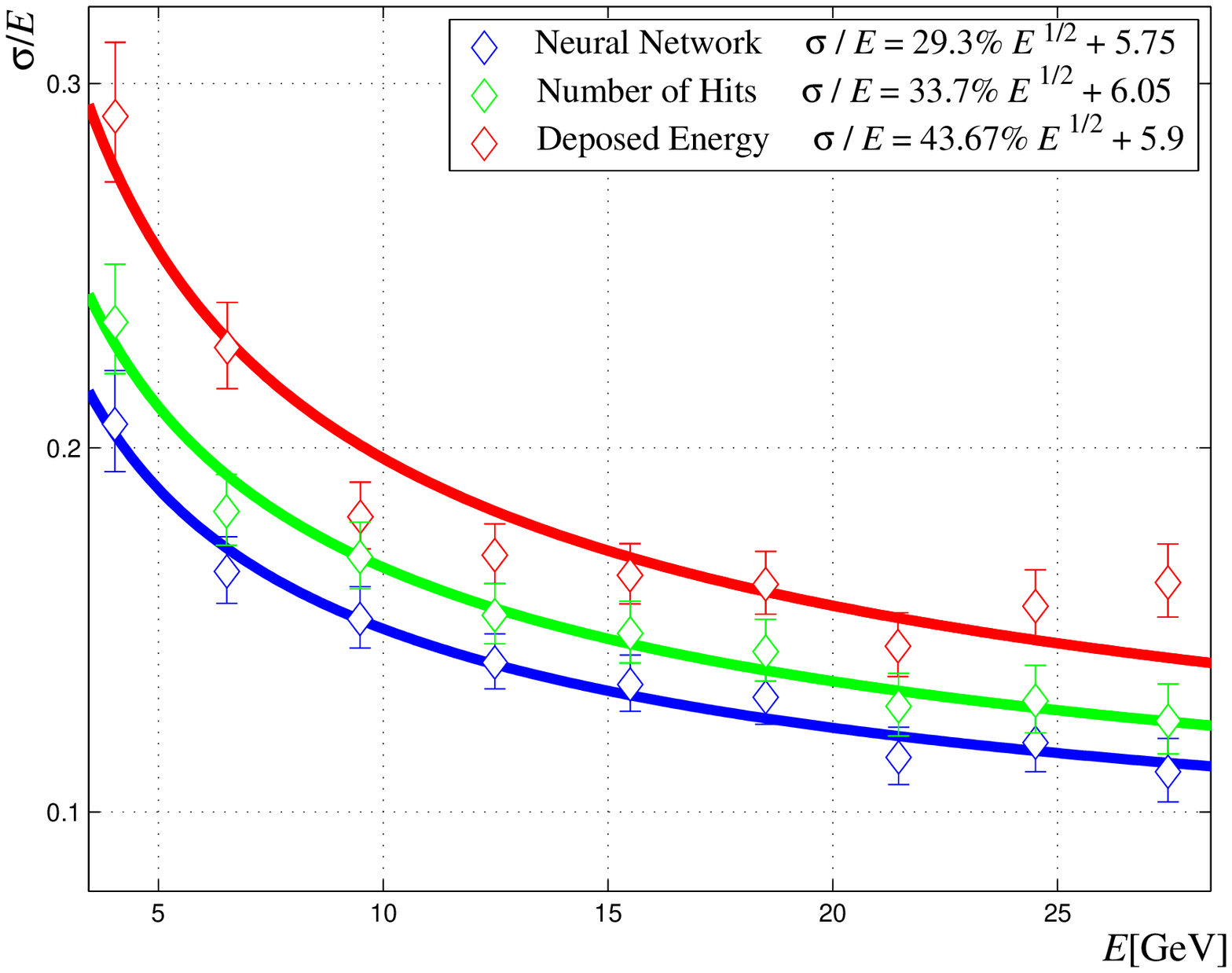}
\caption{The energy resolution using energy deposited in the
  scintillator(red dots), using pad multiplicity(green square) 
 and eventually using more informations from digital pattern (blue triangle).}
\label{fig:domenico}
\end{center}
\end{figure}

\mbox{~}\\[1cm]
We can develop on the more original variant, the digital. Here the spatial
information is preferred to the local energy measurement. The detecting medium
is read in small cells, about 1cm$^2$ which may make some 50 millions cells, 
but purely by yes/no.
This provides an optimal information for separating the showers, hence an
excellent muon identification in particular below 5 GeV where muon chambers 
are useless. This clearly provides also an easy way of handling halo muons.
Then we can wonder about the jet energy resolution. This has been looked at
with a simulation where the detecting medium is a perfect scintillator
structured in 1cm$^2$ cells. Pions of different energies are sent. 
The energy seen in
the cells is summed, the number of cells  is counted and also a neural
net trained for energy resolution out of the cell distribution is used.
Figure \ref{fig:domenico}
  shows the response as a function of energy.

\begin{figure}[htp]
\begin{center}
\includegraphics[width=8.0cm,height=8.0cm]{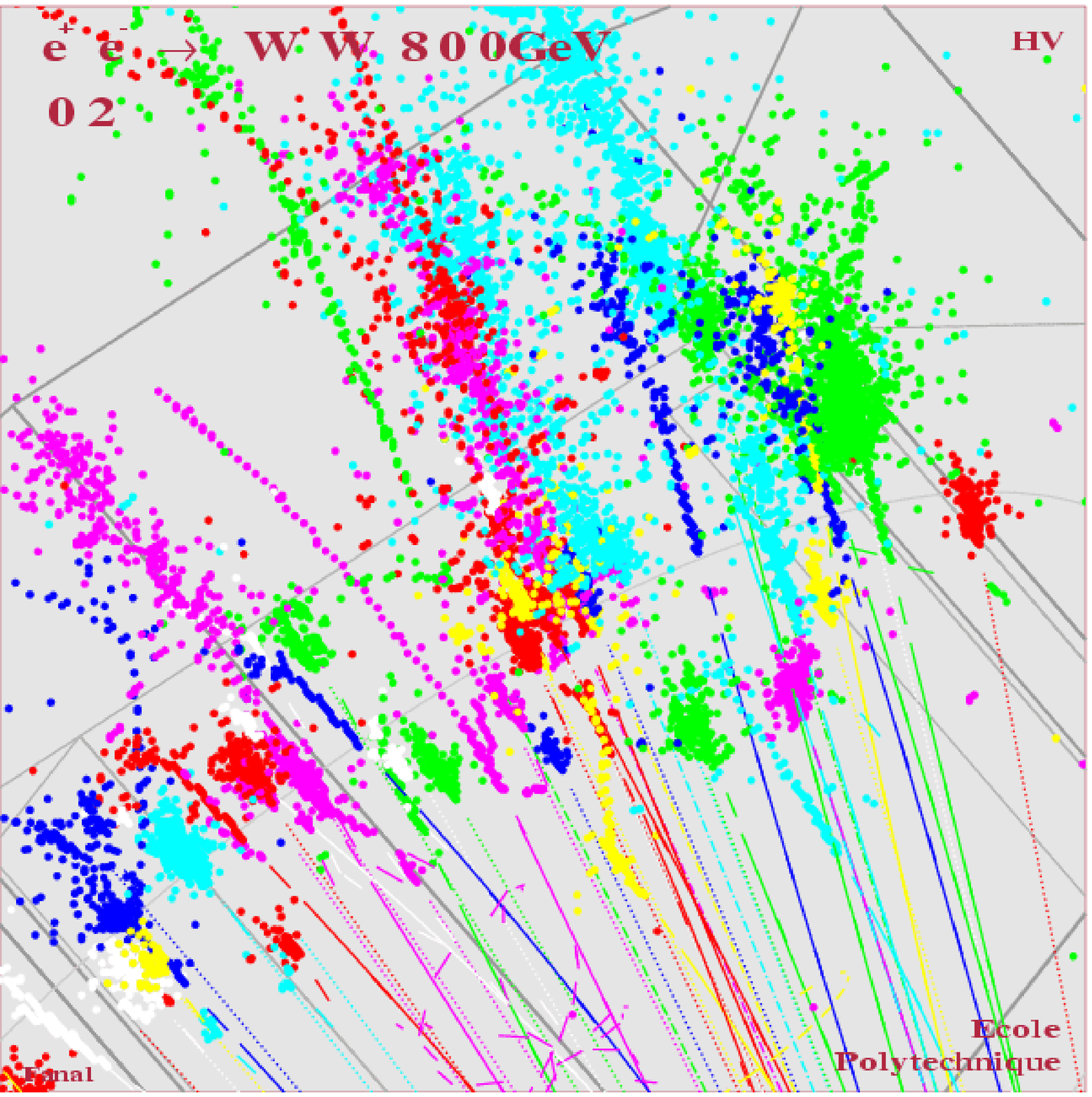}
\caption{ View of W di-jet in \ee$\to$\ww  at $\sqrt{s}$=800  GeV}
\label{fig:ww800}
\end{center}
\end{figure}

 It is clear that,
for this size of cell with that sampling, the counting provides a better
estimate of the energy than summing the cell energies. This a priori surprising
result originates from the suppression of cell fluctuations by the counting.
The jet resolution has been investigated at the Z peak leading to a
resolution of 2.89 GeV as shown on figure [\ref{fig:Zmass}],
 equivalent to an $\alpha$ of 0.3.

The main advantage of the method is nevertheless in the shower separation
and has to demonstrate its power with boosted events. For that purpose, 
reactions like \mbox{\ee$\to$\ww}  at maximum energy (800 GeV) are under study.
This is illustrated in figure \ref{fig:ww800}
 showing a W dijet in the view that best
separates the jet components.

\section{An affordable calorimeter}

\subsection{The W-Si electromagnetic calorimeter}

 The more frequently asked questions  about the ECAL are:\\
 - Is it affordable ?\\
 - By how much the performance is sensitive to the silicon diode
 quality ?\\
 - Does the number of channels (about 35 10$^6$) create 
 specific problems for the calibration,uniformity,etc...\\

\underline{The questions of the silicon diodes quality}\\
  First, to simulate dead wafers,
  a fraction, up to 5\%,  of the  wafers are not used
  in the reconstruction code . It is assumed that there is
 no  geometrical correlation between the dead wafers.
 Because there is a good knowledge of the shower profile,
 (related to the number of layer), 
 an efficient correction can be applied 
 on the reconstructed photon/electron when there
 is dead wafer(s) in the shower path.
  As a consequence,
  the performances of the ECAL
 are alsmot unchanged; i.e. the energy resolution 
 at 1 GeV grows from 10\%/$\sqrt{E}$ (no dead wafer)
 to 10.2\%/$\sqrt{E}$ (5\% of dead wafers). It is 
 important to note that accepting 5\% of dead wafers
 in the production, may increase the industrial yield
 by a factor of 2 \cite{bib:merkin},  which directly changes
 the overall cost of the ECAL by roughly the same amount.\\

\begin{figure}[ht]
\begin{center}
\includegraphics[width=12.0cm,height=8.0cm]{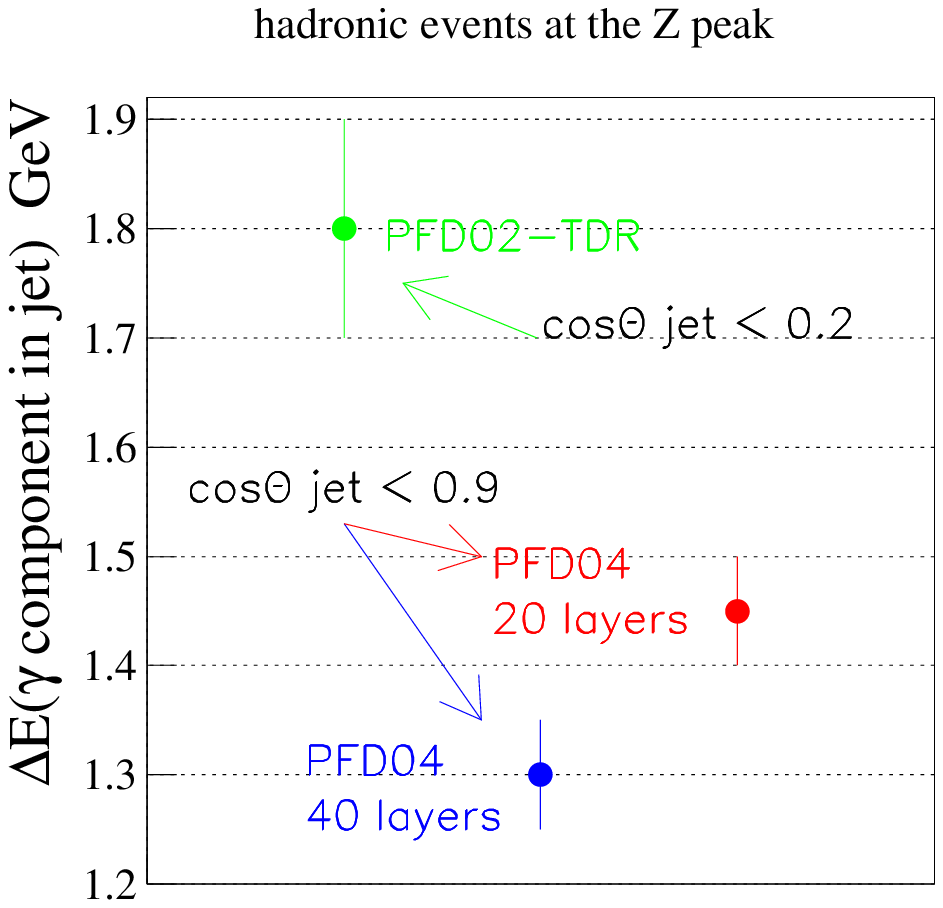}
\caption{ The resolution on the photonic component
 of the jet energy. The value of the TDR (green dots)
 is compared to the new version of the photon reconstruction code
 for  40 (blue dots) and 20 (red dots) layers of silicon.}
\label{fig:40vs20}
\end{center}
\end{figure}

\underline{The total area of silicon}\\
 A second test has been performed by changing the number of
 silicon layers from 40 (TDR) to 20.
 Using the deposited energy, the 
 stochastic term of the energy resolution 
 increases from 10\% to 14\%, but using in addition
 informations like the
 pad multiplicity, the shower profile,etc..., 
 leads to a stochastic term of 12\%.
 To see the impact on jets, the photon reconstruction code
 PFD04 \cite{bib:pfd04} has been adapted to the 20 layers
 geometry. The result, on figure \ref{fig:40vs20},
 given in terms of the energy resolution on the photonic
 component of the jet energy, is an increase of only
  about 10\%.
 Clearly, the case for the 
 silicon-tungsten ECAL is largely unchanged
 when reducing the number of layers from 40 to 20,
 which almost reduces the overall cost of the ECAL
 by a factor two.\\

\underline{A very large number of channels}\\
The total amount of readout pads in the ECAL is about 35 10$^6$.
 This very large number induces strong constraints on the 
 level of the noise in the readout, the electronics cost,etc...
 It can also generate problems with the calibration,
  the non-uniformity,etc... The overall calibration is not a problem
 with the large number of processes involving electron(s),
 where the tracker can provide the momentum of the electron(s).
 It remains the intercalibration and more generally the 
 effect of the non-uniformity of the detector.
  The main sources of  non-uniformity
 are i) the electronics gains, ii) the depletion thickness
 of each diode and iii) the thickness/density
 of the tungsten. It has been pointed out previously \cite{bib:calib}
 that all these sources could be controlled, on-line for the
 point i) and  at the production or construction time
 for the points ii) and iii). The residuals on the correction
 would come from the errors on the measurements.
 Under this condition, the type of distribution we expect
 for the dispersion is gaussian.
 Therefore, to do the study, the pad  responses have been smeared 
 by a gaussian centred at zero with a dispersion
 of 2\% and then of 5\%.
  The resolution  on the final state photons in jet events increases
 by only 70 MeV for a non-uniformity of 5\%,  a value which
  seems largely within reach.
 The main concern remains the coherent noise,
 which will be studied by prototyping.\\

\underline{The question of the cost}\\
\noindent  Summarizing informations on the question of the cost of the ECAL, 
 when compared to the LHC microstrip tracker or pre-shower,
 it can be stated that : 
\mbox{~}\\[0.2cm]
 \noindent $\bullet$ the macroscopic size of the silicon detector (1cm$^2$)
 leads to an easier production;
 \mbox{~}\\[0.cm]
\noindent $\bullet$ the number of masks to produce the silicon matrices
 is about 2 times smaller\cite{bib:merkin};
\mbox{~}\\[0.cm]
 \noindent  $\bullet$ the industrial yield should be 2 times larger;
\mbox{~}\\[0.2cm]
 \noindent In addition,  if needed for the cost,  the number of layers
 could  be decreased by a factor 2, with only a modest degradation
 of the performances on jets. 
 From all these points, the conclusion is that
 {\bf{ the cost is  not a killing  factor for the silicon-tungsten ECAL.}}  

\subsection{The digital hadronic calorimeter}

 When speaking of large area of possible digital chambers in recent 
 experiment,
 the example of Belle detector\cite{bib:belle},
 with 4000 m$^2$ of RPC's  can be taken.
 As a  matter of fact, the technical solutions 
 for the detecting device are under investigation, in 
particular at IHEP-Protvino. The possibility to use RPC's or gas wire chambers
makes the solution quite inexpensive and robust. 
 The yes/no electronics makes even such a large number of channels
 inexpensive. There is still much work to be done.

A full description of the mechanical and electronic solutions
 for the W-Si ECAL as well as for the two 
 options of the HCAL can be found in the TDR \cite{bib:TDR}
 and at the CALICE web site \cite {bib:calice}.

\section{Conclusions}

The electromagnetic calorimeter as a sandwich of tungsten and silicon appears
to be adequate from the point of view of physics, it remains to be fully 
proven technologically. As described in the TDR it may look financially
difficult but compromises are possible which do not harm deeply the
physics performances and reduce very substantially the cost.

For the hadronic part a choice has to be made between the two variants and will
be made on the basis of full simulation and reconstruction as well as some
prototyping. Nevertheless, aside that point, many parameters have to be tuned 
properly like the choice of radiator material.

Mastering the design of a calorimeter for a linear collider needs still
considerable effort on the technical side but also on the software development
and on the physics validation of the performances.

\end{document}